\documentclass[aps,prb,showpacs,superscriptaddress,twocolumn,amssymb,amsfonts]{revtex4}

\usepackage{amsmath}
\usepackage{bm}
\usepackage{graphics} 
\usepackage{dsfont}
\usepackage{subfigure}		
\usepackage{epsfig}

\usepackage[usenames,dvipsnames]{color}			
\definecolor{purple}{rgb}{0.5,0,0.5}

\begin{document}

\title{Weyl Phases in Point-Group Symmetric Superconductors}

\author{Vasudha Shivamoggi}
\affiliation{Department of Physics, University of Illinois, Urbana-Champaign}

\author{Matthew J. Gilbert}
\affiliation{Department of Electrical and Computer Engineering, University of Illinois, Urbana-Champaign}
\affiliation{Micro and Nanotechnology Laboratory, University of Illinois, Urbana-Champaign}

\begin{abstract}
We study superconductivity in a Weyl semimetal with broken time-reversal symmetry and stabilized by a point-group symmetry.  The resulting superconducting phase is characterized by topologically protected bulk nodes and surface states with Fermi arcs.  The topological invariant governing the system is calculated using changes in eigenvalues of the point-group operator along high-symmetry momentum lines.  We show that this invariant is determined by the Fermi surface topology of the Weyl semimetal.  We discuss the effect of surface orientation and $C_4$-breaking strain as possible experimental consequences.
\end{abstract}


\maketitle

\section{Introduction}
\label{sec:Intro}
The discovery of topological phases in non-interacting electron band structures has resulted in a wealth of interesting new systems~\cite{TKNN, HasanKaneRev, QiZhangRev, SchnyderRyuFurusakiLudwig, KitaevClass}.  For materials with fully gapped bands, it is possible to define a topological invariant that determines the number of protected surface states~\cite{LaughlinQH, HalperinQH, FuKaneZ2pump} and the quantized value of response functions such as the Hall conductivity in integer quantum Hall systems~\cite{TKNN, AvronSeilerSimon} or the magnetoelectric effect in strong topological insulators\cite{QiHughesZhanglong, EssinMooreVanderbilt}.  The invariant underlies the precise quantization of these response functions and the robustness of the surface states, since its value does not change as long as the bandgap does not close.  

In addition to fully gapped insulators and superconductors, however, topological phases can occur in gapless systems when the bulk nodal structures are themselves protected~\cite{volovikbook, TurnerVishwanathReview}.  This occurs in Weyl semimetals, where the bulk bands touch at separated pairs of points in the Brillouin zone known as Weyl nodes.  In 3d, these nodes are robust to gap-opening perturbations.  They behave like monopoles of Berry curvature and can only be removed by pairwise annihilation when two nodes with oppositely charged monopoles merge~\cite{Murakami, WanTurnerVishwanathSavrasov}.  The magnitude of the monopole is the topological invariant underlying a pair of Weyl nodes, and is determined by the Berry flux through a closed 2d surface in momentum space that encloses a single node.  Similar to the bulk invariant in gapped topological phases, the bulk Weyl nodes are manifested physically as surface states displaying Fermi arcs and an anomalous Hall effect~\cite{YangLuRan}.  Weyl nodes have been proposed to exist in the A phase~\cite{volovikbook} of $^3$He and recently in several materials including pyrochlore iridates~\cite{WanTurnerVishwanathSavrasov, YangLuRan}, topological insulator-ferromagnet multilayers~\cite{BurkovBalents, HalaszBalents}, and HgCr$_2$Se$_4$~\cite{DaiDoubWeyl, FangMultiweyl}.

In superconductors, nodal topological phases have also been found in certain time-reversal-invariant (TRI) materials whose bulk gap closes at nodal lines or rings in momentum space~\cite{SatoNodal, BeriNodal, RyuSchnyderFlat}.  CePt$_3$Si is one such noncentrosymmetric superconductor in which spin-orbit coupling is proposed to result in a mixed singlet-triplet pairing.  The stability of line nodes in this material has been attributed to a non-trivial topological invariant of a 1d loop in momentum space enclosing the nodal line.  The system has surface states with unusual flat bands in a certain range of momentum determined by the bulk line nodes.

In this work we extend the understanding of topological phases in nodal superconductors to systems with point nodes that break time-reversal symmetry~\cite{SauTewariMajoranaArcs, BalentsWeylSC, ChoWeylSC, AjiWeylSC}.  Specifically we study a superconducting model with double-Weyl nodes (with monopole strength of 2) and protected surface states displaying Fermi arcs.  The model consists of a double-Weyl semimetal stabilized by $C_{4h}$ point-group symmetry, and superconducting pairing terms that preserve the point-group symmetry.  To argue that the nodes and surface states are protected, we derive a topological invariant for the superconductor.  For a Weyl phase the calculation has two steps: find each pair of bulk nodes, and sum the corresponding Berry monopoles.  We derive a simple way to carry out each step using changes in the symmetry eigenvalues along high-symmetry momentum (HSM) lines in the Brillouin zone.

Interestingly, we find that the invariant for the superconducting Weyl nodes is determined by the Fermi surface topology of the normal state.  Fermi surface topology studies have been previously used to characterize fully gapped, TRI superconductors~\cite{SatoGappedLong, QiHughesZhangTriTSC, SatoGappedShort, FuBerg}.  For example, the $\mathbb{Z}_2$ invariant of a gapped, odd-parity TRI superconductor is determined by the number of TRI momenta enclosed by the Fermi surface~\cite{SatoGappedLong, SatoGappedShort, FuBerg}.  We extend these Fermi surface studies to the time-reversal-breaking gapless structure of Weyl phases.  The method is applied to a superconducting system whose pairing vanishes on HSM lines, as the $C_4$-invariant pairings do.  In this case, finding bulk nodes in the superconductor becomes equivalent to finding the intersections of HSM lines with the Fermi surface of the parent material.  Therefore an invariant describing the existence and type of bulk nodes becomes a statement about the Fermi surface topology.

In Sec.~\ref{sec:SCPN}, we argue based on symmetry considerations that Weyl nodes are stable in 3d, time-reversal-breaking superconductors.  We also prove that robust invariants can be constructed from symmetry eigenvalues even in the presence of bulk nodes.  In Sec.~\ref{sec:DoubleWSM}, we present a model for a $C_{4h}$-invariant double-Weyl semimetal.  Pairing terms that respect this symmetry are calculated in Sec.~\ref{sec:pairing}, as well as surface states of the resulting Weyl superconductor.  In Sec.~\ref{sec:invar}, we derive an expression for the topological invariant of the superconducting Weyl phase.  The result is applied to another class of $C_{4h}$-invariant Hamiltonians and generalized to systems with $C_n$ in Sec.~\ref{sec:addcases}.  Finally we give experimental signatures in Sec.~\ref{sec:expsig}.

\section{Superconductors with Point Nodes}
\label{sec:SCPN}

\subsection{Stability of Weyl Nodes in Superconductors}
\label{sec:weylprop}
We argue that Weyl nodes are stable in superconductors that break TRS by considering symmetry classes.  Weyl nodes are pairs of bulk gap-closing points that are topologically protected.  For a 2-band model in a 3d system, the Hamiltonian near the Weyl nodes can be written as $H = v_x p_x \sigma ^x + v_y p_y \sigma ^y \pm v_z p_z \sigma ^z$.  The $\pm$ in front of the last term indicates that the two nodes have opposite chirality.  Any further perturbation commutes with at least one term in the Hamiltonian, and thus can only shift the location of the nodes but cannot open a gap.  The existence of Weyl nodes requires either time-reversal symmetry (TRS) or inversion to be broken, and in this work we consider systems without TRS.

Weyl nodes are topological defects in momentum space.  Similar to vortices in real space, they may annihilate pair-wise but are individually robust.  It is desirable to find a topological invariant underlying a pair of Weyl nodes analogous to the vorticity of a vortex pair.  Because of the presence of bulk nodes, a topological invariant cannot be well-defined over the full 3d Brillouin zone.  Instead it is useful to consider a lower dimensional subset of the Brillouin zone that surrounds the nodal region.  Since the subset is chosen by design to avoid the nodes, the system constrained to the subset is fully gapped.  A topological invariant is defined on this lower dimensional subset so that a non-trivial value of the invariant indicates enclosed nodes.  The robustness of this invariant ensures that the nodes cannot be eliminated except across quantum phase transitions.  Gapless phases are therefore protected by topological invariants of lower dimensional subsets that enclose the nodal region.

\begin{figure}[t]
	\subfigure[]{ \includegraphics[width=0.2\textwidth]{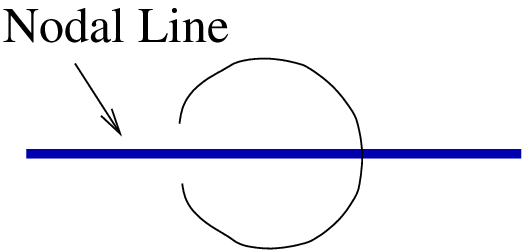}
		\label{nodalline} }
	\subfigure[]{ \includegraphics[width=0.1\textwidth]{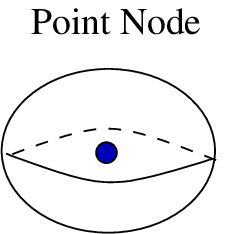}
		\label{nodalpoint} }
	\caption{(a)  (Color online) A nodal line (blue) may be enclosed by a 1d loop (black) in momentum space.  Topologically non-trivial nodal lines are protected by a 1d topological invariant defined on the enclosing loop, and are accompanied by the zero-energy states of the non-trivial 1d system.  (b) A point node (blue) can be enclosed by a 2d surface (black), and is therefore protected by a non-trivial 2d invariant.  The associated surface states are the linearly dispersing mid-gap states corresponding to a topologically non-trivial 2d system.}
	\label{momemsub}
\end{figure}

Topological phases have been classified into a periodic table that predicts the invariants allowed for a given spatial dimension and symmetry class.  We use this classification to first review the topological nature of TRI superconductors with nodal lines, then analyze superconductors with point nodes.  In TRI superconductors, the Hamiltonian on generic subsets of the Brillouin zone breaks both TRS and particle-hole symmetry (PHS).  However the Hamiltonian satisfies a chiral symmetry defined as the product of TRS and PHS, and thus belongs to symmetry class AIII which has a non-trivial integer-valued invariant~\cite{SchnyderRyuFurusakiLudwig} in 1d.  Nodal lines can be surrounded by a 1d loop in momentum-space and are protected by the 1d invariant (Fig.~\ref{nodalline}).  Furthermore the surface states of the superconductor correspond to the non-trivial edge states on the 1d momentum subset, which are zero-energy states.  These flat surface bands may seem surprising in a 3d system, but should be thought of as the edge modes of the 1d topological phase that is protected by the nodal line~\cite{SatoNodal, BeriNodal, RyuSchnyderFlat}.  

Next we turn to superconductors with Weyl nodes and show the stability of those that break time-reversal.  Weyl nodes are point nodes that are enclosed by a 2d subset of the Brillouin zone (Fig.~\ref{nodalpoint}).  Class AIII has no non-trivial 2d topological invariant, so point nodes are unstable in TRI superconductors.  For this reason, breaking inversion while preserving TRS is insufficient for a Weyl superconductor, though of course it is possible that additional symmetries may stabilize it.

For superconductors that break TRS, the Hamiltonian on generic subsets of the Brillouin zone again breaks TRS and PHS.  In this case there is no additional chiral symmetry, so the Hamiltonian on the subsets belongs to class A.  The existence of a 2d topological invariant in class A makes Weyl nodes in TRS-breaking superconductors stable.  The surface states correspond to the edge states of non-trivial 2d systems in class A, which are chiral and have linear dispersion.  We verify this for the model studied in Sec.~\ref{sec:pairing}.  Weyl nodes in superconductors breaking TRS therefore do no need additional symmetries to be stabilized.  
Our argument is in agreement with a previous microscopic calculation~\cite{BalentsWeylSC} showing that superconducting Weyl nodes are unstable in TRS systems but stable when TRS is broken.  In this work  we use certain point-group symmetries to express the corresponding topological invariant in a simple way, though they are not necessary for the existence of the nodes.

\subsection{Role of Fermi Surface Topology}
\label{sec:GenArg}
We now show that symmetry eigenvalues can define a topological invariant even in the presence of generic bulk gap closings.  Consider a two-band model 
\begin{equation}
h_0 = d_1 \sigma ^x + d_2 \sigma ^y + d_3 \sigma ^z
\label{toyham}
\end{equation}
where $d_m$ are functions of momentum $\mathbf{k}$, and $\sigma ^m$ are Pauli matrices in a basis of spin, orbital, or spin-orbit-coupled degrees of freedom.  A point group operator $\eta$ for a $C_n$, or $n$-fold, rotation about the $z$- axis maps $\mathbf{k}$ to $U\mathbf{k} = (k_x \cos \frac{2\pi}{n} - k_y \sin \frac{2\pi}{n}, k_x \sin \frac{2\pi}{n} + k_y \cos \frac{2\pi}{n}, k_z)$.  In this work we consider a $C_{4h}$ symmetry consisting of a combined 4-fold rotation about the $z$-axis and mirror reflection about the $xy$-plane, which take $(k_x, k_y, k_z)$ to $(k_y, -k_x, -k_z)$.  A system that preserves this symmetry obeys
\begin{equation}
\eta h_0(\mathbf{k}) \eta ^{\dagger} = h_0(U \mathbf{k}).
\label{SymmInvar}
\end{equation}
Without loss of generality, consider the case where the point-group operator is represented by $\sigma ^z$.  This leads to the following contstraints on the coefficients of the Pauli matrices:
\begin{subequations}
\begin{equation}
d_1(U \mathbf{k}) = -d_1(\mathbf{k})~,~d_2(U \mathbf{k}) = -d_2(\mathbf{k})
\label{SymCoefOdd}
\end{equation}
\begin{equation}
d_3(U \mathbf{k}) = d_3(\mathbf{k})
\label{SymCoefEven}
\end{equation}
\end{subequations}
We define high-symmetry momentum (HSM) points $\mathbf{\Gamma _a}$ as points in momentum space left invariant under the symmetry operator: $U \mathbf{\Gamma _a} = \mathbf{\Gamma _a}$.  Evaluating Eq.~\ref{SymCoefOdd} at HSM gives $d_{1,2}(\mathbf{\Gamma _a}) = - d_{1,2}(\mathbf{\Gamma _a})$, which implies that $d_1$ and $d_2$ vanish there.  At these special points, the Hamiltonian takes the form
\begin{equation}
h_0(\mathbf{\Gamma _a}) = d_3(\mathbf{\Gamma _a}) \sigma ^z~,
\label{HSham}
\end{equation}
with energy eigenvalues $E(\mathbf{\Gamma _a}) = \pm |d_3(\mathbf{\Gamma _a})|$.  The Hamiltonian at HSM commutes with the symmetry operator, so each energy band can be characterized by eigenvalues of the symmetry operator, $\nu_i$:
\begin{equation}
\nu (\Gamma _a) = -\text{sgn} ~d_3(\mathbf{\Gamma _a})~.
\label{HSeig}
\end{equation}
The eigenvalue $\nu (\Gamma _a)$ changes sign only when $d_3 (\mathbf{\Gamma _a})$ goes to 0, \textit{ie}, when the energy gap closes at one of the HSM.  The product of symmetry eigenvalues, $\chi = \prod_a \nu (\Gamma _a)$, can be used to define phases as regions of constant $\chi$ separated by gap closings at $\mathbf{\Gamma _a}$.  Since the $\nu (\Gamma _a)$ are defined only at HSM $\Gamma _a$, they do not change when the gap closes at non-high-symmetry points.  Therefore $\chi$ is a robust invariant even in the presence of bulk gap closings, as long as they occur away from $\Gamma _a$.

\begin{figure}[t]
	\subfigure[]{ \includegraphics[width=0.2\textwidth]{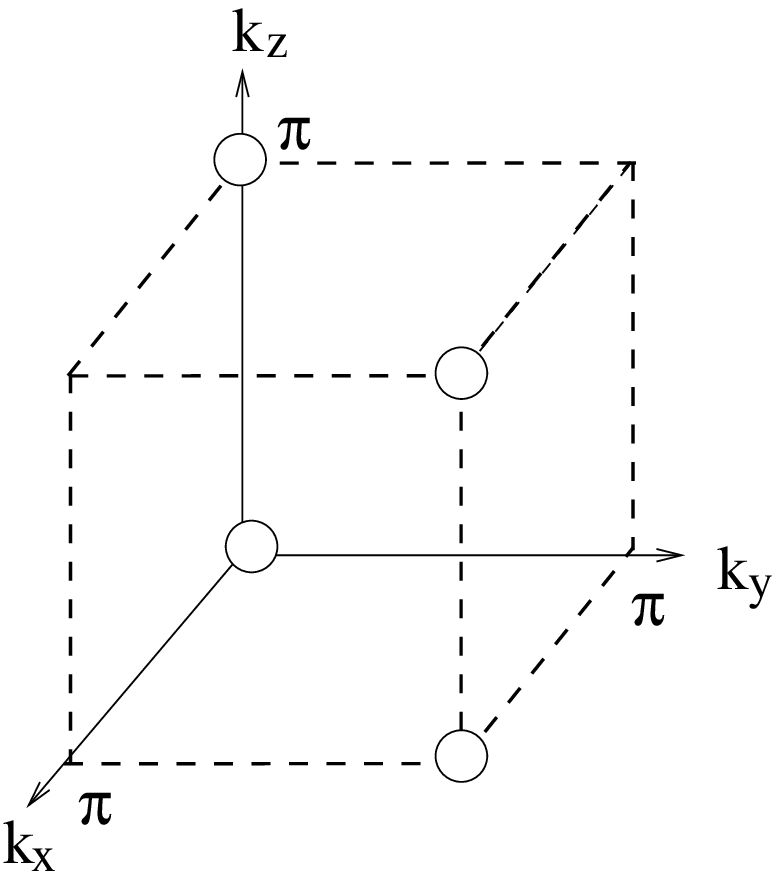}
		\label{samptr} }
	\subfigure[]{ \includegraphics[width=0.2\textwidth]{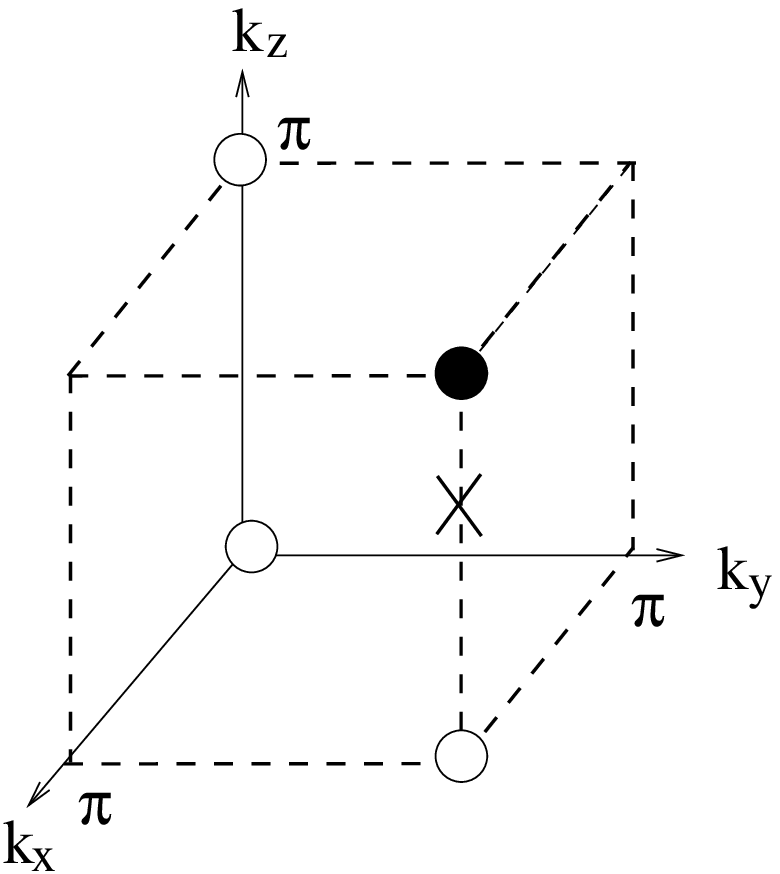}
		\label{samtop} }
	\caption{Typical $C_4$ eigenvalue configurations for (a) an insulator and (b) a Weyl semimetal.  The white (black) circles represent $\nu = +1~(-1)$.  The invariant $\chi _{\mathbf{k_{\perp}}}$ is the ratio of $\nu$ at $k_z = 0$ to $\nu$ at $k_z = \pi$ for fixed $\mathbf{k_{\perp}}$.  A value of $\chi | _{\mathbf{k_{\perp}}} \neq 1$ indicates a bulk node (represented here by a cross) along the line $\mathbf{k_{\perp}}$. For the semimetal, $\chi _{\pi} = -1$, indicating a node along the $(k_x, k_y) = (\pi, \pi)$ line.}
	\label{eigconfig}
\end{figure}

In addition to defining a robust invariant for nodal systems, the symmetry eigenvalues can be used to detect and classify the nodes themselves.  Consider rotation-invariant momentum lines, for example $\mathbf{k_{\perp}} = (k_x, k_y) = (0, 0)$ or $(\pi, \pi)$ for $C_4$-invariant systems.  Along a fixed $\mathbf{k_{\perp}}$, unequal values of $\nu$ at $k_z = 0$ and $k_z = \pi$ indicate that the gap has closed at some value of $k_z$ in between.  Formally this difference is expressed as 
\begin{equation}
\chi | _{\mathbf{k_{\perp}}} = \prod _n \frac{ \nu _n (k_{\perp}, k_{\perp}, 0)}{ \nu _n (k_{\perp}, k_{\perp}, \pi)}~.
\label{chi_diff}
\end{equation}
This quantity is an invariant within each phase, and a value different from unity indicates the presence of a node (Fig.~\ref{eigconfig}).  In Sec.~\ref{sec:invar}, we will show that the specific value of $\chi _{\mathbf{k_{\perp}}}$ determines the Berry monopole strength of the nodes, for example 2 for double-Weyl nodes.

This argument also applies to a superconductor with protected bulk nodes.  Here we show that examining symmetry eigenvalues at fixed $\mathbf{k_{\perp}}$ is equivalent in certain superconductors to studying the Fermi surface topology.  Consider a point-group symmetric Weyl semimetal, and add pairing that respects the symmetry.  The Hamiltonian of the superconductor is
\begin{equation}
H = H_0(\mathbf{k}) + \mu \tau ^z + \Delta (\mathbf{k})~,
\label{BDGgen}
\end{equation}
where $H_0(\mathbf{k})$ corresponds to the single-particle Hamiltonian $h_0$ from Eq.~\ref{toyham} after Bogoliubov-de-Gennes (BdG) doubling, $\mu$ is the chemical potential, $\Delta (\mathbf{k})$ is the pairing matrix, and $\tau ^i$ are the Pauli matrices in the electron and hole basis.  As before, both $d_1 (\Gamma _a)$ and $d_2 (\Gamma _a)$ are required to vanish by the underlying point-group symmetry.  Additionally we consider pairing terms that vanish on rotationally-invariant momentum lines.  $C_{4h}$-invariant pairings satisfy this constraint, as will be shown in Sec.~\ref{sec:pairing}, however the following argument applies to any pairing term that goes to zero along high-symmetry lines.  This leaves two non-zero terms in the Hamiltonian at $\Gamma _a$: $d_3 (\Gamma _a)$ and $\mu$.  The chemical potential commutes with the spin and orbital degrees of freedom represented by $\sigma ^m$.  At HSM, the energy bands are
\begin{equation}
E(\Gamma _a) = \pm \left[d_3(\Gamma _a) \pm \mu \right]~.
\label{BdGenergyHSM}
\end{equation}   
As a result of the simplified form of $H$ at the HSM, we can again define an invariant in terms of symmetry eigenvalues
\begin{equation}
\nu '(\Gamma _a) = -\text{sgn}\left[d_3 (\Gamma _a)\pm \mu\right]~.
\label{HSeig_SC}
\end{equation}
The eigenvalues change sign when the bulk gap closes at HSM, analogous to the semimetal.  The key difference is that the gap-closing occurs at $d_3(\Gamma _a) = 0$ in the semimetal and $d_3(\Gamma _a) = \pm \mu$ in the superconductor.  For superconductors whose pairing vanishes on rotationally-invariant momemtum lines, the latter condition has a simple physical meaning in terms of the normal (non-superconducting) material.  It marks the intersection of the parent Fermi surface with the rotationally-invariant momentum lines.  Therefore bulk nodes in this type of superconductor are equivalent to intersections of HSM lines with the parent Fermi surface.  Analogous to Eq.~\ref{chi_diff}, an invariant $\chi | _{\mathbf{k_{\perp}} }$ is now written in terms of eigenvalues of the superconducting system, Eq.~\ref{BDGgen}.  As in the semimetal case, $\chi | _{\mathbf{k_{\perp}} }$ is different from unity in the presence of bulk nodes, and the exact value of $\chi | _{\mathbf{k_{\perp}} }$ determines the Berry monopole associated with the bulk nodes.  The relation between $\chi | _{\mathbf{k_{\perp}} }$ and the bulk invariant describing Weyl nodes is the focus of Sec.~\ref{sec:invar}.

This argument does not depend on the specific form of the pairing term, other than that it vanishes at HSM.  In this sense the result is similar to previous work studying the effect of odd-parity pairing on topological insulators, where the topological invariant is determined by Fermi surface topology~\cite{SatoGappedLong, QiHughesZhangTriTSC, SatoGappedShort, FuBerg}.  For example, the strong 3d index is equal to the number of enclosed TRI momenta modulo 2~\cite{SatoGappedLong, SatoGappedShort, FuBerg}.  Weak indices are found using intersections of the Fermi surface with HSM, which for TRS are planes at $k_i = \pi$~\cite{SatoGappedLong, SatoGappedShort}.  This is analogous to the invariants we construct using Fermi surface intesections with HSM lines to study Weyl nodes protected by a weak topological insulator phase.

\section{Double-Weyl Semimetal}
\label{sec:DoubleWSM}
We first discuss the properties of a double-Weyl semimetal with $C_{4h}$-symmetry before considering pairing.  We use the basis states $| \frac{3}{2} \frac{3}{2}\rangle$ and $|S, -\frac{1}{2}\rangle$, or equivalently $|c_{p_x + ip_y, \uparrow}\rangle$ and $|c_{s, \downarrow}\rangle$.  These basis states are both eigenstates of angular momentum with $J_z = \frac{3}{2}$ and $-\frac{1}{2}$ respectively.  Physically this is due to strong spin-orbit coupling, which couples lattice rotations to rotations in spin space.  Under a $C_4$ rotation about the $z$-axis, the orbitals transform as $p_x + ip_y \rightarrow -i(p_x + ip_y)$ and $s \rightarrow s$.  Additionally the spin $\uparrow$ and $\downarrow$ pick up factors of $e^{\mp i\pi/4}$ respectively.  Putting both phases together, the $C_{4h}$ operator in this basis has the form $e^{-\frac{3\pi i}{4}} \sigma ^zU(\pi/2)$, where $U$ maps $(k_x, k_y, k_z)$ to $(k_y, -k_x, -k_z)$.  There are four HSM defined by $k_x = k_y = 0$ or $\pi$ and $k_z = 0$ or $\pi$.

Using this basis, we consider a 2-band Hamiltonian following a derivation for HgCr$_2$Se$_4$~\cite{DaiDoubWeyl}:
\begin{align}
\nonumber h_0 = &\left( \cos k_x - \cos k_y\right) \sigma ^x + \sin k_x \sin k_y \sigma ^y \\
&+ \left( m - \cos k_x - \cos k_y - \cos k_z\right) \sigma ^z
\label{3dHam}
\end{align}
This model preserves a combined $C_4$ and $M_z$ symmetry but breaks time-reversal symmetry, making it a 3d class A system.  As expected when the symmetry operator proportional is to $\sigma ^z$, the coefficients in front of $\sigma ^x$ and $\sigma ^y$ vanish at HSM.  Eq.~\ref{3dHam} has bulk gap closings in a certain range of $m$: when $1< m< 3$, the nodes are at $(k_x, k_y, k_z) = (0, 0, \pm k_0)$, where $k_0 = \cos ^{-1}(m-2)$.  Similarly for $-3 < m < -1$, the nodes are at $(\pi, \pi, \pi \pm k_{\pi})$ for $k_{\pi} = \cos ^{-1}(m+2)$.  $1 < |m| < 3$ therefore corresponds to a phase with two distinct nodes that merge at the phase boundaries.

To see that this is a Weyl phase, we expand the Hamiltonian near the two nodes using $\vec{k} = (\delta k_x, \delta k_y, k_0 + \delta k_z)$ for $m>0$ (the $m < 0$ case proceeds similarly):
\begin{equation}
h_0 \approx -\frac{1}{2} \delta k_- ^2 \sigma ^+ - \frac{1}{2} \delta k_+ ^2 \sigma ^- \pm \sqrt{1 - (m-2)^2}\delta k_z \sigma ^z
\label{WeylExpansion}
\end{equation}
where $k_{\pm} \equiv k_x \pm ik_y$ and $\sigma ^{\pm} \equiv \sigma ^x \pm i \sigma ^y$.  The $\pm$ in front of the last term in Eq.~\ref{WeylExpansion} indicates that the two nodes have opposite chirality.  The Hamiltonian near the gapless points is quadratic in $k_x$ and $k_y$.  These two facts indicate that for $1< |m| < 3$, the model has double-Weyl nodes.

A robust invariant for the Weyl semimetal cannot be defined over the full 3d Brillouin zone because of the presence of nodes at two values of $k_z$.  However fixing $k_z$ at a value away from the Weyl nodes results in a fully gapped 2d system for which a well-defined invariant does exist.  As $k_z$ is swept through various values, a change in the 2d invariant indicates a Weyl node.  The magnitude of the change is equal to the strength of the associated Berry monopole.  In the Weyl phase (eg, $-3 < m < -1$), the 2d invariant is $C = 2$ in between the Weyl nodes $\pi - k_{\pi} < k_z < \pi + k_{\pi}$, and $C = 0$ otherwise.  The change in Chern number, $|\Delta C| = 2$, is the topological invariant protecting the double-Weyl phase.

\begin{figure}[t]
	\subfigure[]{ \includegraphics[width=0.35\textwidth]{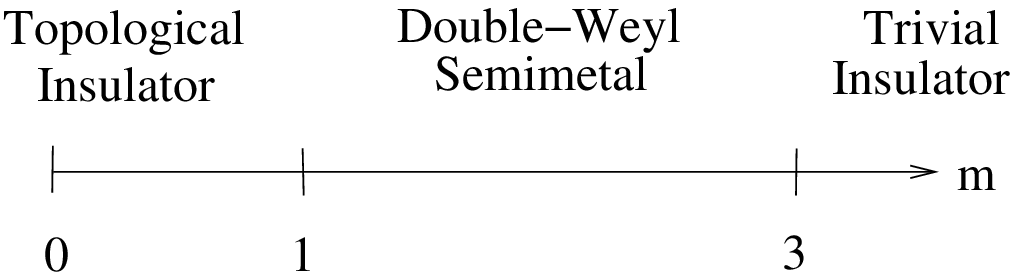}
		\label{doubpd} }
	\subfigure[]{ \includegraphics[width=0.35\textwidth]{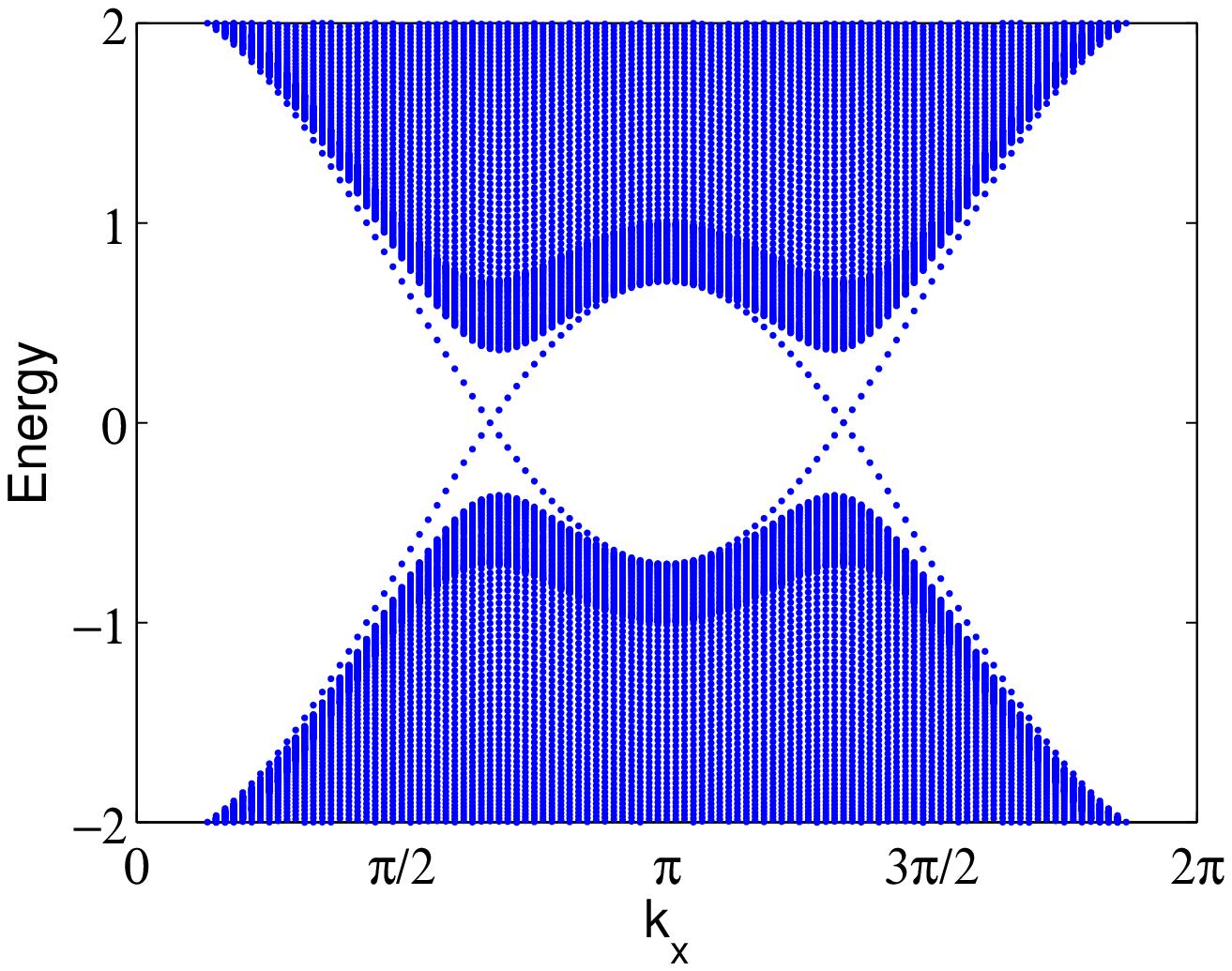}
		\label{WSMsurfacespec} }
	\caption{(a) Phase diagram of the model Hamiltonian Eq.~\ref{3dHam} for $m \geq 0$ (the phases are symmetric about $m = 0$).   For the insulating phases, the Chern number is well-defined over the full Brillouin zone and takes the values $C = 2$ and $C = 0$ for the topological and trivial phases respectively.  For the double-Weyl semimetal phase, the invariant is the difference in Chern number on 2d momentum space slices separated by a Weyl node, $\Delta C = 2$. (b) Energy bands of the double-Weyl semimetal with open boundaries in the $y$-direction at fixed $k_z = \pi$ and $m = -2$.  For this  value of $k_z$ there are two chiral states per surface, indicating a topological invariant equal to 2. }
	\label{SMSurfaceSpec}
\end{figure}

Apart from a few exceptions~\cite{HughesProdanBernevig, TurnerZhangMongVishwanath}, a non-zero value of a topological invariant is accompanied by robust surface states.  We now show that 2d slices at fixed $k_z$ where $C = 2$ have two chiral mid-gap states per surface, while those in the $C = 0$ range of $k_z$ have no surface states.  This momentum-dependant manifestation of the bulk-boundary correspondence results in the unusual Fermi arcs on Weyl surfaces.  We consider surfaces parallel to the $xz-$plane and calculate the spectrum of states localized in the $y-$direction as a function of $k_x$ for various values of fixed $k_z$.  A previously derived method~\cite{MongShivamoggi} expresses the existence of surface states as well as their dispersion in terms of parameters of the bulk Hamiltonian $h_0$ (Eq.~\ref{3dHam}).  Using this method, it can be shown that the surface states exist in the following range of $k_z$:

\begin{equation}
\vert \frac{1}{2} (m-\cos k_z)\rvert < 1 ~.
\label{edgecondition}
\end{equation}
This indicates that surface states exist when $|m| < 3$.  Note that the range of mass $m$ resulting in a Weyl semimetal, $1 < |m| < 3$, is only a subset of the parameter space allowing surface states.  The model therefore has more than one type of topological phase.  The other phase, at $|m| <1$, is a weak topological insulating phase (Fig.~\ref{doubpd}).  

\begin{figure}[t]
\begin{center}
\includegraphics[width=0.35\textwidth]{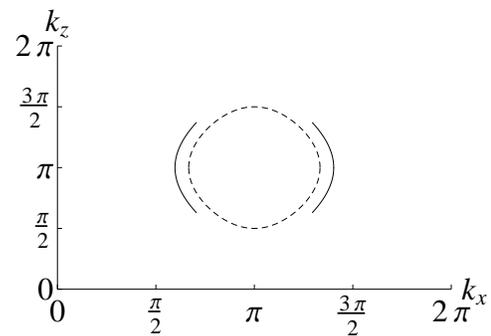}
\end{center}
\caption{\label{WSMfermiarcs} Fermi arcs for $m = -2$.  The dashed line corresponds to a $C_4$-invariant system, which has two Fermi arcs beginning and ending at the two double-Weyl nodes.  The two nodes split into four double-Weyl nodes when a $C_4$-breaking term is added, and the Fermi arcs (solid lines) terminate at these four nodes.  In this case, it is clear that the Fermi surface is comprised of open arcs.}
\end{figure}

For a value of $k_z$ satisfying Eq.~\ref{edgecondition}, the energy of the surface states can be found analytically~\cite{MongShivamoggi} to be
\begin{equation}
E(k_x) = \pm \left[ \cos k_x - \frac{1}{2} (m-\cos k_z)\right]~.
\label{Eedge}
\end{equation}
Fig.~\ref{WSMsurfacespec} shows both bulk and surface bands of the Weyl semimetal at $k_z = \pi$.  There are two chiral states per surface, with degeneracies at $k_x = \pi \pm \cos ^{-1}[(m+1)/2]$.  As $k_z$ changes, the surface state degeneracies shift according to Eq.~\ref{Eedge} until finally they merge at the Weyl nodes, $k_z = \pi \pm k_{\pi}$.  As $k_z$ is increased further into the region where $C = 0$, there are no surface states at all.  Fig.~\ref{WSMfermiarcs} shows a plot of the Fermi surface at $E = 0$ for $m = -2$.  The dashed lines are two Fermi arcs terminating at the double-Weyl nodes.  The bulk-boundary correspondence can therefore be formulated in Weyl phases as an equivalence between the number of robust Fermi arcs and the Berry monopole strength of the bulk Weyl node.  The Fermi arcs arise because the surface states only exist for the range of $k_z$ between the two Weyl nodes.  The Weyl semimetal owes its existence to the topological insulator in this region.  As $m$ approaches $\pm 1$ or $\pm 3$, the two nodes approach one another and the range of $k_z$ with $C = 2$ shrinks.  It vanishes entirely at the phase boundaries when the Weyl nodes merge, and further changes in $m$ result in a fully gapped phase.  

The double-Weyl nodes can be thought of as two single-Weyl nodes with $\Delta C = 1$ that have merged and are kept together by $C_4$-symmetry~\cite{FangMultiweyl}.  If $C_4$ is broken, for example by adding the term $a_0 \sigma ^x$ to Eq.~\ref{3dHam}, each double-Weyl node splits into two single-Weyl nodes.  Note that the Chern number $C$ is still 2 for $k_z$ between the Weyl nodes, indicating that the topological properties of the surface states are robust to $C_4$-breaking terms.  The surface spectra shift so that the Fermi arcs terminate at the four single-Weyl nodes, seen as the solid lines in Fig.~\ref{WSMfermiarcs}.  Now it is clear that they are two disconnected arcs.  The role of the point-group symmetry is to stablize multi-Weyl nodes in this model, but the topological phase itself does not depend on this symmetry for its existence.  As we showed in Sec.~\ref{sec:weylprop}, a 3d Hamiltonian in class A can support Weyl nodes without the need for additional symmetries.  The motivation for considering point-group-symmetric models in the present work is that the symmetry makes evaluating the invariant easier.

\section{$C_4$-Invariant Pairing}
\label{sec:pairing}
We turn to a superconductor with Eq.~\ref{3dHam} as its kinetic term and pairing that preserves $C_{4h}$.  To be particle-hole symmetric, the Hamiltonian $H$ must satisfy
\begin{equation}
\Xi H(k) \Xi ^{\dagger} = -H^T(-k)~,
\label{PHSconstraint}
\end{equation}
where $\Xi$ is the PHS operator.  For a general Hamiltonian of the form
\begin{equation}
\nonumber H = \left( \begin{array} {cc}
h_e (\mathbf{k}) & \delta (\mathbf{k})\\
\delta ^{\dagger}(\mathbf{k}) & h_h (\mathbf{k})
\end{array} \right)
\end{equation}
and $\Xi = \tau ^x$, PHS requires
\begin{subequations}
\begin{equation}
h_h(\mathbf{k}) = -h_e^T(-\mathbf{k})~,
\label{PHSh}
\end{equation}
\begin{equation}
\delta (\mathbf{k}) = -\delta ^T(-\mathbf{k})~.
\label{PHSd}
\end{equation}
\end{subequations}
The kinetic part of the Hamitonian becomes
\begin{align}
\nonumber H_0 = &\left( \cos k_x - \cos k_y\right) \sigma ^x \tau ^z + \sin k_x \sin k_y \sigma ^y \\
&+ \left( m - \cos k_x - \cos k_y - \cos k_z\right) \sigma ^z \tau ^z + \mu \tau ^z~.
\label{3dHamBdG}
\end{align}
$C_{4h}$ symmetry is expressed as
\begin{equation}
\eta H(k_x, k_y, -k_z) \eta ^{\dagger} = H (k_y, -k_x, -k_z)~.
\label{C4constraint}
\end{equation}
In the full electron-hole basis, the $C_{4h}$ operator has the form
\begin{align}
\nonumber \eta &= \left( \begin{array} {cccc}
e^{-\frac{3\pi i}{4}}&&&\\
& e^{\frac{\pi i}{4}}&&\\
&&e^{\frac{3\pi i}{4}}&\\
&&&e^{-\frac{\pi i}{4}}
\end{array} \right)\\
&= \frac{-1}{\sqrt{2}}U(\frac{\pi}{2}) \left(\sigma ^z + i\sigma ^z \tau ^z \right)
\label{fullC4}
\end{align}

We find four pairing terms that satisfy both PHS and $C_{4h}$, Eqs.~\ref{PHSd} and \ref{C4constraint}:
\begin{subequations}
\begin{equation}
\Delta _1 = \Delta _0 \left(\sin k_x \tau ^x - \sin k_y \tau ^y \right)~,
\label{C4Pairing1}
\end{equation}
\begin{equation}
\Delta _2 = \Delta _0 \left(\sin k_x \sigma ^x \tau ^x - \sin k_y \sigma ^x\tau ^y \right)~,
\label{C4Pairing2}
\end{equation}
\begin{equation}
\Delta _3 = \Delta _0 \sin k_z \left(\sin k_x \sigma ^y \tau ^x - \sin k_y \sigma ^y\tau ^y \right)~,
\label{C4Pairing3}
\end{equation}
\begin{equation}
\Delta _4 = \Delta _0 \left(\sin k_x \sigma ^z \tau ^x - \sin k_y \sigma ^z\tau ^y \right)~.
\label{C4Pairing4}
\end{equation}
\end{subequations}
The full Hamiltonian is $H = H_0 + \Delta _i$.  The Hamiltonian for the $C_{4h}$-invariant superconductor is a four-band model with nodes shifted from the single-particle Weyl nodes to $(0, 0, \pm \cos ^{-1}(m-2 \pm \mu))$ or $(\pi, \pi, \pi \pm \cos ^{-1}(m+2 \pm \mu))$.  Since all of the $C_{4h}$-invariaint pairings vanish along $C_4$-invariant momentum lines, they cannot open a gap at the superconducting nodes.  

As a first check on the protected nature of the superconducting nodes, we show that they are robust to gap-opening in the same way as the single-particle nodes.  The protection of the latter arises  from the observation that a node of a two-band model written in terms of 3 Pauli matrices requires tuning 3 independent parameters.  In 3d, the momenta provide a means to use up all possible mass terms, allowing for robust nodes.  A four-band model can be written in terms of five mutually anticommuting $4\times 4$ matrices, suggesting that Weyl behavior is possible in d = 5.  However the nodes in our d = 3 model are protected as a result of the form of the kinetic part of the Hamiltonian, Eq.~\ref{3dHamBdG}.  It can be shown that no matrix anticommutes with all three terms of $H_0$.  Any term added to $H_0$ commutes with at least one term, which at most shifts the Weyl node.  

\begin{figure}[t]
	\subfigure[]{ \includegraphics[width=0.35\textwidth]{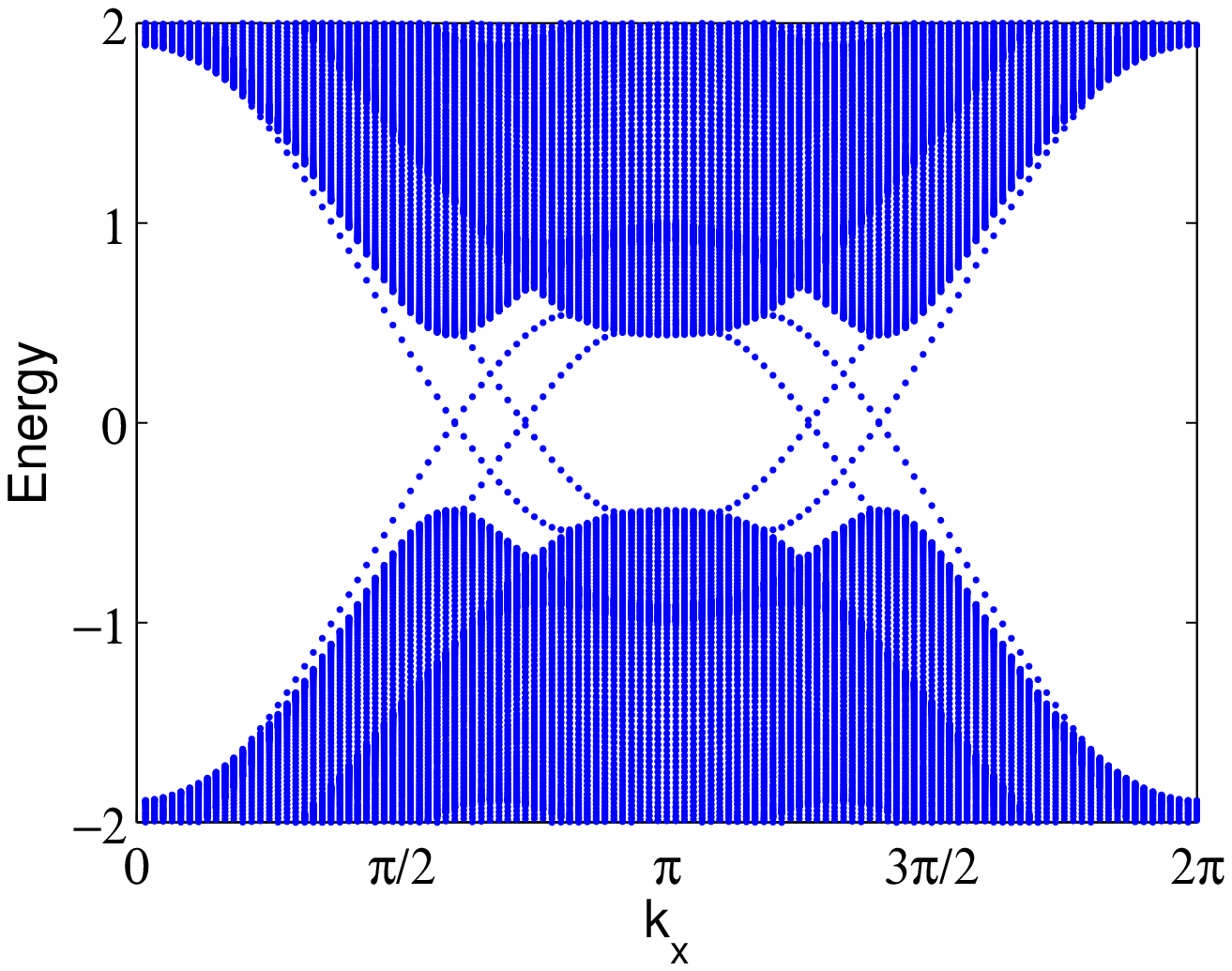}
		\label{SCsplitsurface} }
	\subfigure[]{ \includegraphics[width=0.2\textwidth]{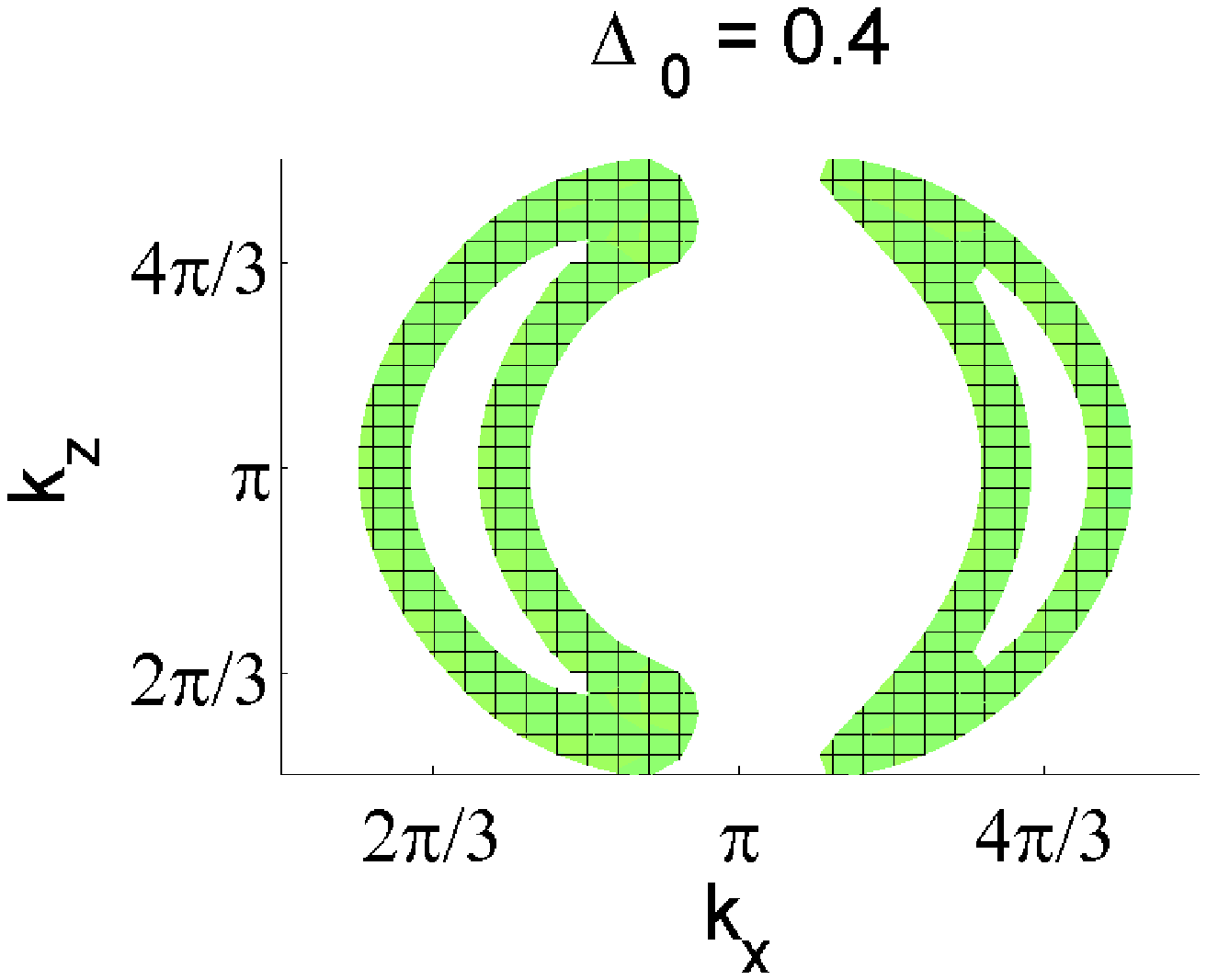}
		\label{MajoArcs4} }
		\subfigure[]{ \includegraphics[width=0.2\textwidth]{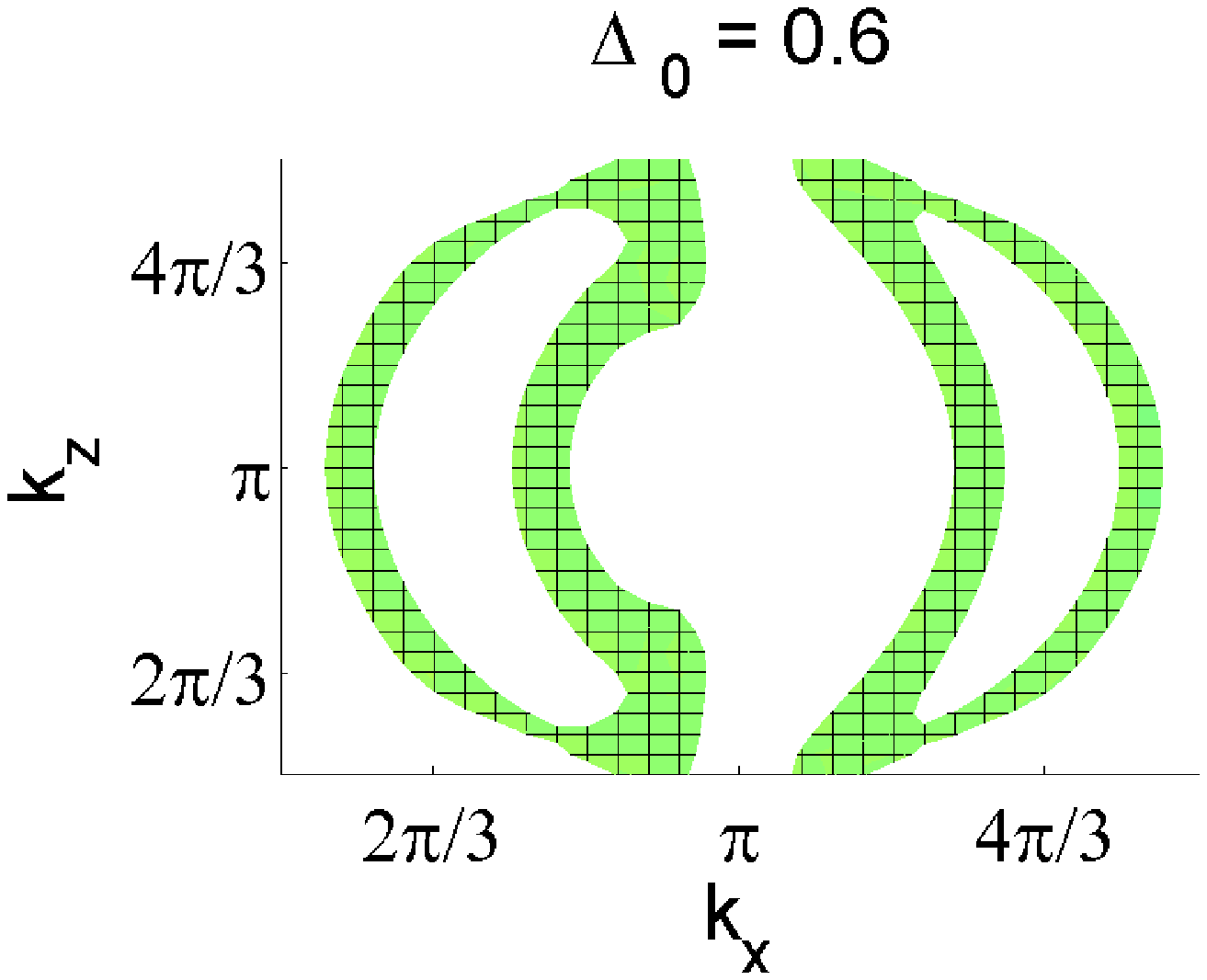}
		\label{MajoArcs6} }
	\caption{Surface spectrum and Fermi arcs with pairing term $\Delta _2$ at $m = -2$.   (a) Surface bands at fixed $k_z = \pi$ for $\Delta _0 = 0.4$.  Comparison with Fig.~\ref{WSMsurfacespec} shows that pairing causes each single-particle surface state to split into two.  The number of surface states, in this case 4, is protected by the bulk topological invariant of the superconducting system.  (b) There are four Fermi arcs, in contrast to the two arcs in the absence of pairing (Fig.~\ref{WSMfermiarcs}).  The splitting of each pair of surface states increases with $\Delta _0$.  (c) shows Fermi arcs at $\Delta _0 = 0.6$.}
	\label{SCSurfaceSpec}
\end{figure}

The superconducting Weyl nodes are associated with surface states, and we present the spectrum with pairing $\Delta _2$ (Eq.~\ref{C4Pairing2}) as a representative example.  Fig.~\ref{SCsplitsurface} shows the energy eigenvalues for $H = H_0 + \Delta _2$ with open boundaries in the $y$-direction.  The model parameters are the same as those in Fig.~\ref{WSMsurfacespec}: $m = -2$, and $k_z = \pi$.  Comparing with Fig.~\ref{WSMsurfacespec} shows that pairing splits each of the two surface states into two, by an amount that grows with $\Delta _0$.  The spectra for $\Delta _3$ and $\Delta _4$ are similar, while $\Delta _1$ results in two pairs of doubly degenerate surface states.  The splitting (or lack thereof) can be understood as the response of the semimetal surface states to perturbations.  Each pairing matrix commutes with at least one term in $H_0$ and shifts the energy eigenvalues accordingly.  For the pairing in our example, $\left[\Delta _2, \sigma ^z \tau ^z \right] = 0$, and the corresponding bulk energies are written as
\begin{align}
\nonumber E_2^2 = &(\cos k_x - \cos k_y)^2 +(\sin k_x \sin k_y )^2  \\
&+ \left[m - \sum_i \cos k_i \pm \Delta _0 \sqrt{\sin ^2 k_x + \sin ^2 k_y} \pm \mu\right]^2 ~.
\label{SCeig2}
\end{align}
$\Delta _2$ enters the energy spectrum as two momentum-dependent perturbations of opposite sign, $\pm \Delta _0 \sqrt{\sin ^2 k_x + \sin ^2 k_y}$, to the $\sigma ^z \tau ^z$ term in $H_0$.  A constant perturbation $a_0 \sigma ^z \tau ^z$ shifts the surface states, in opposite directions for opposite signs of $a_0$.  The effect of $\Delta _2$ is to shift the surface states in opposite directions and cause a splitting proportional to $\Delta _0$, as seen in Figs.~\ref{MajoArcs4}-\ref{MajoArcs6}.  The same explanation applies to the effect of $\Delta _3$ and $\Delta _4$.  However $\Delta _1$ enters the bulk energy spectrum as a perturbation to the $\sigma ^y$ term.  It can be shown that adding a constant perturbation $a_0 \sigma ^y$ leaves the surface states unshifted, and this qualitatively explains the two-fold degeneracy of the surface states with $\Delta _1$.

\section{Calculation of the Superconducting Invariant}
\label{sec:invar}
The superconducting Weyl nodes and surface states studied in the previous section are protected by a bulk topological invariant.  We now calculate this invariant using changes in symmetry eigenvalues along HSM lines.  We first derive a formula for the invariant for a semimetal in Sec.~\ref{sec:C4InvariantSingP} and then extend the analysis to a superconductor in Sec.~\ref{sec:C4InvariantBdG}.
\subsection{Semimetal Invariant}
\label{sec:C4InvariantSingP}
For a 3d system with $C_4$ symmetry in the $xy-$plane, the Chern number $C$ at a fixed value of $k_z$ is given up to a gauge by
\begin{equation}
e^{\frac{i\pi}{2} C} = \prod _{n \in occ} \xi _n(0, 0, k_z) \xi _n (\pi, \pi, k_z) \zeta _n (0, \pi, k_z)~,
\label{C42bandChern}
\end{equation}
where $\xi$ and $\zeta$ are $C_4$ and $C_2$ eigenvalues respectively~\cite{FangMultiweyl}.  Eq.~\ref{C42bandChern} provides a simple way of computing the 2d invariant for any fixed value of $k_z$ away from a node.  Unequal values of $C$ for two different $k_z$ indicate Weyl nodes.  According to the argument presented in Sec.~\ref{sec:GenArg}, the nodes can be detected directly via changes in symmetry eigenvalues along HSM lines, \textit{ie} at fixed $\mathbf{k_{\perp}} = (k_x, k_y)$ rather than $k_z$, as in Eq.~\ref{C42bandChern}.  We now prove that these changes determine the type of Weyl node as well.  The advantage of this alternative formula is that it brings to light the role of the parent material's Fermi surface topology in the superconducting case.  From Eq.~\ref{C42bandChern}, the change in Chern number $\Delta C$ between $k_z = 0$ and $k_z = \pi$ is
\begin{equation}
e^{\frac{i\pi}{2}\Delta C} = \prod _{n \in occ} \frac{\xi _n(0, 0, 0)\xi _n(\pi, \pi, 0)\zeta _n(0, \pi, 0)}{\xi _n(0, 0, \pi)\xi _n(\pi, \pi, \pi)\zeta _n(0, \pi, \pi)}~.
\label{ChernC4Ratio}
\end{equation}
The eigenvalues of the $C_4$ and $C_2$ operators take the form $e^{i\pi s/2}$ and $e^{i\pi r}$ respectively, where $s$ and $r$ are integers.  Two values of $\xi (\zeta)$ can be compared using their ratio, $e^{i\pi (s_a - s_2)/2} \equiv e^{i\theta} (e^{i\phi})$.  For convenience we define $\theta _{k_{\perp}}$ as the phase difference of two eigenvalues $\xi$ at the same $k_x = k_y = k_{\perp}$.  Substituting these forms into Eq.~\ref{ChernC4Ratio} gives 
\begin{equation}
\nonumber e^{\frac{i\pi}{2}\Delta C} = e^{i\theta _0} e^{i\theta _{\pi}} e^{i\phi}~.
\end{equation}
The expression for $\Delta C$ is
\begin{equation}
\Delta C = \frac{1}{\pi/2}\left( \phi + \sum_{k_{\perp}} \theta _{k_{\perp}} \right)~.
\label{ChernC4Sum}
\end{equation}
The invariant for a $C_4$-invariant Weyl semimetal is thus determined by the relative phase of the symmetry eigenvalues across the node.  Eq.~\ref{ChernC4Sum} computes the invariant in two steps.  (1) Nodes are detected via changes in $\xi$ or $\zeta$.  Along lines of fixed $k_x$ and $k_y$, the eigenvalues $\xi$ and $\zeta$ can only change when the bulk gap closes at some intermediate $k_z$.  Therefore a non-zero value of any of the phase changes $\theta _{k_{\perp}}$ or $\phi$ indicate a bulk node.  (2) The Berry monopole strength is the sum of phase changes in $\xi$ and $\zeta$.  Eq.~\ref{ChernC4Sum} shows how $C_4$-symmetry permits both single- and double-Weyl nodes.  In models where $\xi$ changes between $+1$ and $-1$, $\theta _{k_\perp} = \pi$ and $\Delta C = 2$ (double-Weyl).  On the other hand single-Weyl nodes may occur when $\xi$ changes between $+1$ and $i$, resulting in $\theta _{k_{\perp}} = \pi/2$ and $\Delta C = 1$.

We apply this result to the double-Weyl model studied in Sec.~\ref{sec:DoubleWSM}.  For the Hamiltonian $h_0$ in Eq.~\ref{3dHam}, the $C_4$ operator was found to be proportional to $\sigma ^z$.  $C_2 = C_4 ^2$ is then proportional to the identity, so the two bands have the same inversion eigenvalue.  In this section, $\zeta(0, \pi, k_z)$ can therefore be set to 1 and dropped from the expressions.  Define high-symmetry momenta (HSM) of the $C_{4h}$ operation as $\Gamma _a = (k_{\perp}, k_{\perp}, k_{z_a})$, where $k_{\perp}$ and $k_{z_a}$ can be 0 or $\pi$ independently.  Evaluating the Hamiltonian at HSM gives $h_0(\Gamma _0) = (m - 2\cos k_{\perp} - \cos k_{z_a})\sigma ^z$, with $C_4$ eigenvalues $\xi(\Gamma _a) = -\text{sgn}(m - 2\cos k_{\perp} - \cos k_{z_a})$.  We use this expression to calculate the invariant in Eq.~\ref{ChernC4Sum} at representative values of $m$.  

At $m = 2$,
\begin{align}
\nonumber k_{\perp} = 0: \xi (0, 0, 0) &= 1~,~\xi(0, 0, \pi) = -1~\Rightarrow \theta _0 = \pi ~,\\
\nonumber k_{\perp} = \pi: \xi (\pi, \pi, \pi) &= -1~,~\xi(\pi, \pi, \pi) = -1~\Rightarrow \theta _{\pi} = 0 ~.
\end{align}
Substituting $\theta _0$ and $\theta _{\pi}$ into Eq.~\ref{ChernC4Sum} gives $\Delta C = 2$, indicating a double-Weyl semimetal at $m = 2$.  The non-zero value of $\theta _0$ additionally shows that the nodes occur along the $\mathbf{k_{\perp}} = (0, 0)$ line.  Repeating this for $m = 4$ gives
\begin{align}
\nonumber k_{\perp} = 0: \xi (0, 0, 0) &= -1~,~\xi(0, 0, \pi) = -1~\Rightarrow \theta _0 = 0 ~,\\
\nonumber k_{\perp} = \pi: \xi (\pi, \pi, \pi) &= -1~,~\xi(\pi, \pi, \pi) = -1~\Rightarrow \theta _{\pi} = 0 ~.
\end{align}
Now $\Delta C = 0$, indicating an insulator at $m = 4$.  We can determine the type of insulator by evaluating Eq.~\ref{C42bandChern} at any value of $k_z$.  All such 2d slices at $m = 4$ have $C = 0$, so this phase is a trivial insulator.  Repeating this exercise at $m = 0$ shows an insulator with $C = 2$ at all values of $k_z$, corresponding to a topological insulator.  The phase diagram in $m$ (Fig.~\ref{doubpd}) shows that the Weyl semimetal interpolates between trivial and topological insulator phases.

\subsection{Superconducting Invariant}
\label{sec:C4InvariantBdG}
The invariant for a $C_{4h}$-symmetric superconductor with Weyl nodes can be similarly written in terms of symmetry eigenvalues.  To evaluate the invariant for Eq.~\ref{3dHamBdG} (independent of the choice of $\Delta _i$), we must find an appropriate symmetry operator~\cite{SatoGappedLong, FuBerg}.  The operator $\tilde{\eta}$ is chosen to satisfy four constraints: (1) it is a symmetry operator of the Hamiltonian, $\tilde{\eta} H(\mathbf{k}) \tilde{\eta}^{\dagger} = H(U\mathbf{k})$; (2) its eigenvalues $\tilde{\xi}$ are defined only at HSM of the $C_{4h}$ symmetry; (3) $\tilde{\xi}$ takes only the values of the single-particle $C_{4h}$ operator; and (4) When the gap closes, $\tilde{\xi}$ changes by the same phase as in the single-particle case.  Eg, $\tilde{\xi}$ must switch sign for the basis used in Eqs.~\ref{3dHam} and~\ref{3dHamBdG}.  The $C_{4h}$ operator in the BdG basis, Eq.~\ref{fullC4}, satisfies constraints (1-3) but not (4).  Of the other natural choices, $\sigma ^z$ and $\sigma ^z \tau ^z$, $\sigma ^z \tau ^z$ is the only one that meets all the requirements.  

Once the appropriate symmetry operator is identified, calculating the invariant is simply a 4-band version of the method derived in Sec.~\ref{sec:C4InvariantSingP}.  The superconducting invariant is the sum of any phase changes in symmetry eigenvalues along HSM lines:
\begin{equation}
\Delta C _{SC} = \frac{1}{\pi/2}\left( \phi _s + \sum_{k_{\perp}} \theta _{s, k_{\perp}} \right)~,
\label{ChernC4SumBdG}
\end{equation}
where $\phi _s$ and $\theta _{s, k_{\perp}}$ are defined for the superconducting system as $\phi$ and $\theta _{k_{\perp}}$ are for the semimetal.  As before, a non-zero value of either $\phi _s$ or $\theta _{s, k_{\perp}}$ indicates a bulk gap-closing.  Additionally it is equivalent to intersections of HSM lines with the parent material Fermi surface in a superconductor with pairing that vanishes along those lines.  Weyl phases in a $C_{4h}$-symmetric superconductor can therefore be characterized by the number of Fermi surface intersections of the parent material.  Using $\tilde{\eta} = \sigma ^z \tau ^z$ gives $\Delta C _{SC} = 4$, in agreement with the four chiral surface states seen in Fig.~\ref{SCsplitsurface}.

\section{Additional Cases}
\label{sec:addcases}
Up to this point, we have studied a 2-band model with $C_4$ eigenvalues of opposite sign.  The other possibility~\cite{YangLuRan} is for a relative phase of $\pi/2$, so that the $C_4$ and $C_2$ operators have the form
\begin{equation}
C_4 = \left( \begin{array}{cc}
1&\\
&i
\end{array} \right)~~,~~P = C_4 ^2 = \left( \begin{array}{cc}
1&\\
&-1
\end{array} \right)~.
\label{C4caseii}
\end{equation}
An example of a 2-band model invariant under $C_{4h}$ in this basis is 
\begin{align}
\nonumber h_0 = &~\alpha \sin k_y \sigma ^x + \alpha \sin k_x \sigma ^y \\
&+ \left( m - \cos k_x - \cos k_y - \cos k_z\right) \sigma ^z~.
\label{3dHamii}
\end{align}
$h_0$ has nodes at $(0, 0, \pm \cos ^{-1}(m-2))$ for $1 < m < 3$ and $(\pi, \pi, \pi \pm \cos ^{-1}(m+2))$ for $-3 < m < -1$.  Additionally, there are two sets of nodes at $(0, \pi, \pm \cos ^{-1}m)$ and $(\pi, 0, \pm \cos ^{-1}m)$ for $|m| < 1$.  Near the nodes for $1 < m < 3$, $h_0$ has the form
\begin{equation}
H \approx \alpha \delta k_y \sigma ^x + \alpha \delta k_x  \sigma ^y \pm \sqrt{1 - (m-2)^2}\delta k_z \sigma ^z
\label{WeylExpanii}
\end{equation}
indicating that these are single Weyl nodes.  The other nodes show similar behavior.

\begin{figure}[t]
\begin{center}
\includegraphics[width=0.35\textwidth]{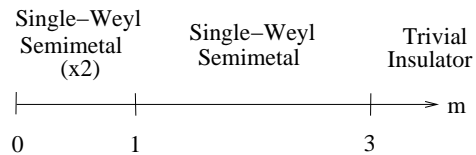}
\end{center}
\caption{\label{singpd} Phase diagram of the model Hamiltonian Eq.~\ref{3dHamii} for $m \geq 0$ (the diagram is again symmetric about $m = 0$).  Both nodal phases contain single-Weyl nodes.  The $0 < m < 1$ phase has two single-Weyl pairs and therefore an invariant $\Delta C  = 2$.  For $1 < m < 3$, $\Delta C = 1$, consistent with one pair of single-Weyl nodes.  The trivial nature of the insulating phase is confirmed by calculating the Chern number $C = 0$ according to Eq.~\ref{3dHam}.}
\end{figure}

We can again use Eq.~\ref{C42bandChern} to calculate invariants and derive a phase diagram for this model.  In this case, the inversion eigenvalue must be included explicitly since it varies between the two bands.  For $m = 2$, this procedure shows that $\xi (0, 0, k_{z_{a}})$ changes from 1 to $i$, leading to $\theta _0 = \pi /2$ and $\Delta C = 1$.  The latter indicates a single-Weyl pair.  At $m = 0$ where $\theta _0 = \theta _{\pi} = 0$ and $\phi = \pi$.  These phase changes result in $\Delta C = 2$.  By $C_4$-symmetry a single-Weyl pair at $\mathbf{k} = (\pi, 0, \cos ^{-1} m)$ must be accompanied by another pair at $\mathbf{k} = (0, \pi, \cos ^{-1} m)$, leading to an invariant of 2.  A phase diagram for this model is shown in Fig.~\ref{singpd}.

It is possible to find 4 $C_{4h}$-invariant pairings in this basis and show that they vanish along $\mathbf{k_{\perp}} = (0, 0)$ and $(\pi, \pi)$.  The invariant for the resulting Weyl superconductor can be found by choosing an appropriate symmetry operator and summing phase changes of its eigenvalues across Fermi surface intersections.

The simple form of the invariant, Eq.~\ref{ChernC4Sum}, also makes it possible to generalize to systems with $C_n$-symmetry.  For the case of $C_2$, the Chern number is determined by the product of inversion eigenvalues~\cite{HughesProdanBernevig, TurnerZhangMongVishwanath}.  In terms of phase changes $\phi$, the invariant is
\begin{equation}
\Delta C = \frac{1}{\pi}\sum_{k_2} \phi _{k_2}~,
\label{ChernC2Sum}
\end{equation}
where $k_2$ is invariant under $C_2$ (\textit{eg}, $k_x = 0$ and $k_y = \pi$).  Similarly the Chern number of a $C_3$-symmetric system is given by the product of $C_3$ eigenvalues~\cite{FangMultiweyl}.  For $\alpha$ defined as the change in phase of a $C_3$ eigenvalue, the invariant is
\begin{equation}
\Delta C = \frac{1}{2\pi /3}\sum_{k_3} \alpha _{k_3}~,
\label{ChernC3Sum}
\end{equation}
where $k_3$ are the $C_3-$invariant momentum lines.  Finally the Chern number of a $C_6$-symmetric model is the product of a $C_2$, $C_3$, and $C_6$ eigenvalue~\cite{FangMultiweyl}.  It can be written as 
\begin{equation}
\Delta C = \frac{1}{\pi /3}\left(\phi + \alpha + \beta \right)~,
\label{ChernC6Sum}
\end{equation}
where $\beta$ is the phase change of the $C_6$ eigenvalue.

\section{Experimental Signatures}
\label{sec:expsig}
Since the bulk density of states vanishes at the Weyl nodes, the surface states dominate the density of states near $E = 0$.  Angle-resolved photoemission spectroscopy or STS measurements are a direct way to probe the surface Fermi arcs.  The superconductor with double-Weyl nodes has two pairs of arcs that split with $\Delta _0$ as seen in Figs.~\ref{MajoArcs4}-\ref{MajoArcs6}, and this may provide a way to measure the strength of the pairing in the material.  

An additional probe is the effect of strain that breaks $C_4$-symmetry.  Under a strain that breaks $C_4$-symmetry, each double-Weyl node splits into two single-Weyl nodes (Fig.~\ref{WSMfermiarcs}).  In the superconducting case, the Fermi arcs split into four open pieces.  The localized states also show a strong dependence on the orientation of the surface.  The surface states shown in Figs.~\ref{WSMsurfacespec} and~\ref{SCsplitsurface} are localized in the $y$-direction.  $C_4$-symmetry implies the same behavior for [100] surfaces.  However no robust surface states exist on [001] surfaces since there are no topologically non-trivial 2d slices that include the $k_z$-axis. 


\section{Conclusion}
\label{concl}
We have studied topologically protected point nodes in superconductors that break time-reversal and preserve $C_4$ symmetry.  An argument based on symmetry classes shows that 3d time-reversal-breaking superconductors can support Weyl nodes with no additional symmetries, unlike TRI superconductors.  We have used a model with $C_{4h}$ symmetry that stabilizes double-Weyl nodes and allows the topological invariant of the system to be easily computed.  The addition of pairing that obeys the same lattice symmetry results in a superconductor with two pairs of double-Weyl nodes.  The topological invariant protecting this phase is calculated from changes in symmetry eigenvalues along HSM lines.  Additionally, the expressions are generalized for systems satisfying $C_n$ symmetry.

The advantage of comparing eigenvalues along HSM lines is that it provides insight into the role of the parent Fermi surface.  We have shown that when the pairing terms vanish along $C_4$-invariant momentum lines, point nodes in the superconductor are equivalent to intersections of the HSM lines with the Fermi surface of the \textit{parent material}.  The role of Fermi surface topology in determining topological invariants has been studied in gapped TRI superconductors with odd pairing.  We have shown that it is similarly pivotal in gapless phases where the invariant protects the nodal structure itself.  This result highlights the importance of Fermi surface topology in the search for unconventional or odd-parity pairing.

\acknowledgements
The authors thank Chen Fang for helpful discussion.  This work was supported by the Department of Energy, Division of Materials Science under Award No. DE-FG02-07ER46453 and the AFOSR under grant FA9550-10-1-0459 (V. S) and the ONR under grant N0014-11-1-0728 (M. J. G.).  


\end{document}